\documentclass[epj,final,fleqn]{svjour}

\usepackage{graphics}
\usepackage{color}
\usepackage{amssymb}
\usepackage{amsmath}
\usepackage{pifont}

\begin{document}
\title{The $^{16}$O$(e,e'pp)^{14}$C reaction to discrete final states}
\author{J. Ryckebusch \and W. Van Nespen}

\institute{Department of Subatomic and Radiation Physics, \protect\\
Ghent University, Proeftuinstraat 86, B-9000 Gent, Belgium
\email{jan.ryckebusch@UGent.be}} \date{\today}

\abstract{ 
Differential cross sections for exclusive
$^{16}$O$(e,e'pp)^{14}$C processes are computed within a
distorted-wave framework which includes central short-range
correlations and intermediate $\Delta ^+$ excitations.  The cross
sections are compared to high-resolution data from the MAMI facility
at Mainz for a central energy and momentum transfer of
$<\omega>$=215~MeV and $<q>$=316~MeV respectively.  A fair agreement
between the numerical calculations and data is reached when using
spectroscopic information extracted from a $^{15}$N$(d,^{3}He)^{14}$C
experiment.  The comparison between the calculations and the data
provides additional evidence that short-range correlations exclusively
affect nucleon pairs with a small center-of-mass momentum
residing in a relative $S$ state.  
}
 
\maketitle

%
%
%



\section{Introduction}
The presence of short-range correlations (SRC) in nuclei is intimately
connected to the finite extension of the nucleons.  Over the years, a
variety of experiments have been hunting for signatures of these
SRC. Major progress has been made with experiments involving electrons
as initial probe.  More in particular, the study of double-coincidence
$A(e,e'p)$ processes has considerably improved our knowledge about the
dynamics of protons embedded in a nuclear medium. The major share of
the $A(e,e'p)$ experiments has been conducted in quasi-elastic
kinematics, whereby the experimental parameters are adjusted so as to
provide a relation between the measured differential cross sections
and the momentum distribution of the hit proton.  The momentum
distribution provides us with knowledge about the probability that a
nucleon in a well-defined orbit has a given value of momentum.  After
correcting for final-state interaction (FSI) effects, the
momentum-dependence of the extracted momentum distributions nicely
reproduced the predictions of non-relativistic and relativistic
nuclear mean-field approaches up to momenta approaching the Fermi
momentum $k_F \approx 250$~MeV.  At the same time, from the absolute
magnitude of the measured cross sections it could be inferred that
when integrating over the mean-field part of the momentum
distributions one ends up with a value which is about 70\% of what
could be expected on the basis of the amount of protons populating the
target nucleus.  This depletion is attributed to the presence of
sizable short- and long-range correlations in atomic nuclei of which
one believes that they exhibit a rather complex radial, spin, isospin
and tensor behavior.  An economical way of parameterizing
nucleon-nucleon correlations is the introduction of correlation
operators with a strength determined by radial-dependent correlation
functions \cite{pieper92}. The determination of these correlation
functions is pivotal in the study of many correlated systems and the
nucleus represents no exception in this matter.  The nuclear central
correlation function, which corresponds with the unity correlation
operator, is believed to have similar characteristics as the
two-particle correlation functions (or, radial distribution functions)
of molecules in liquids \cite{hecht}.  Indeed, when moving with a
nucleon in the nucleus, its finite extension will induce a reduced
probability of finding another nucleon over distances of the order of
its radius $R_p$ and an increased probability at distances slightly
larger than $R_p$.  The radial distribution function for molecules in
liquids shows a similar fluctuating behaviour and can be understood
through the molecule-molecule repulsion extending over distances of
the size of a molecule. Usually, this repulsion is modeled with the
aid of a Lennard-Jones potential \cite{hecht}.

Experimentally determining the correlation function turns out to be
challenging.  It is expected that triple-coincidence
reactions of the $A(e,e'pp)$ type could improve our knowledge about the
dynamics of nucleon pairs and help in mapping the radial dependence of
the central correlation function.  Pioneering experimental work was done at the
AMPS electron accelerator in Amsterdam \cite{zondervan,kester1995,gerco1998}.
Building on this experience, in the final years of operation of this
facility, high-quality data could be collected for the
$^3$He$(e,e'pp)n$ \cite{groep2002} and $^{16}$O$(e,e'pp)^{14}$C process
\cite{starink}.  Complementary measurements on the $^{12}$C
\cite{raoul} and $^{16}$O$(e,e'pp)$ \cite{guenther2000,kahrau1999}
reaction have been performed at the 850~MeV electron accelerator in
Mainz.  These two-proton knockout measurements have sparked off a lot
of theoretical activity of which some of the more recent ones include
the work reported in Refs.~\cite{schawmb2003,kadrev2003,anguiano2003}.
An ambitious two-proton knockout calculation aiming at consistently
computing the long- and short-range correlations in $^{16}$O, in
combination with a treatment of final-state interaction effects, has
been presented in Ref.~\cite{giusti1998}.  Recently, this model has
been extended to include the mutual interactions between the two
ejected protons \cite{schawmb2003}.
 
\begin{figure}
\begin{center}
\resizebox{0.50\textwidth}{!}{\includegraphics{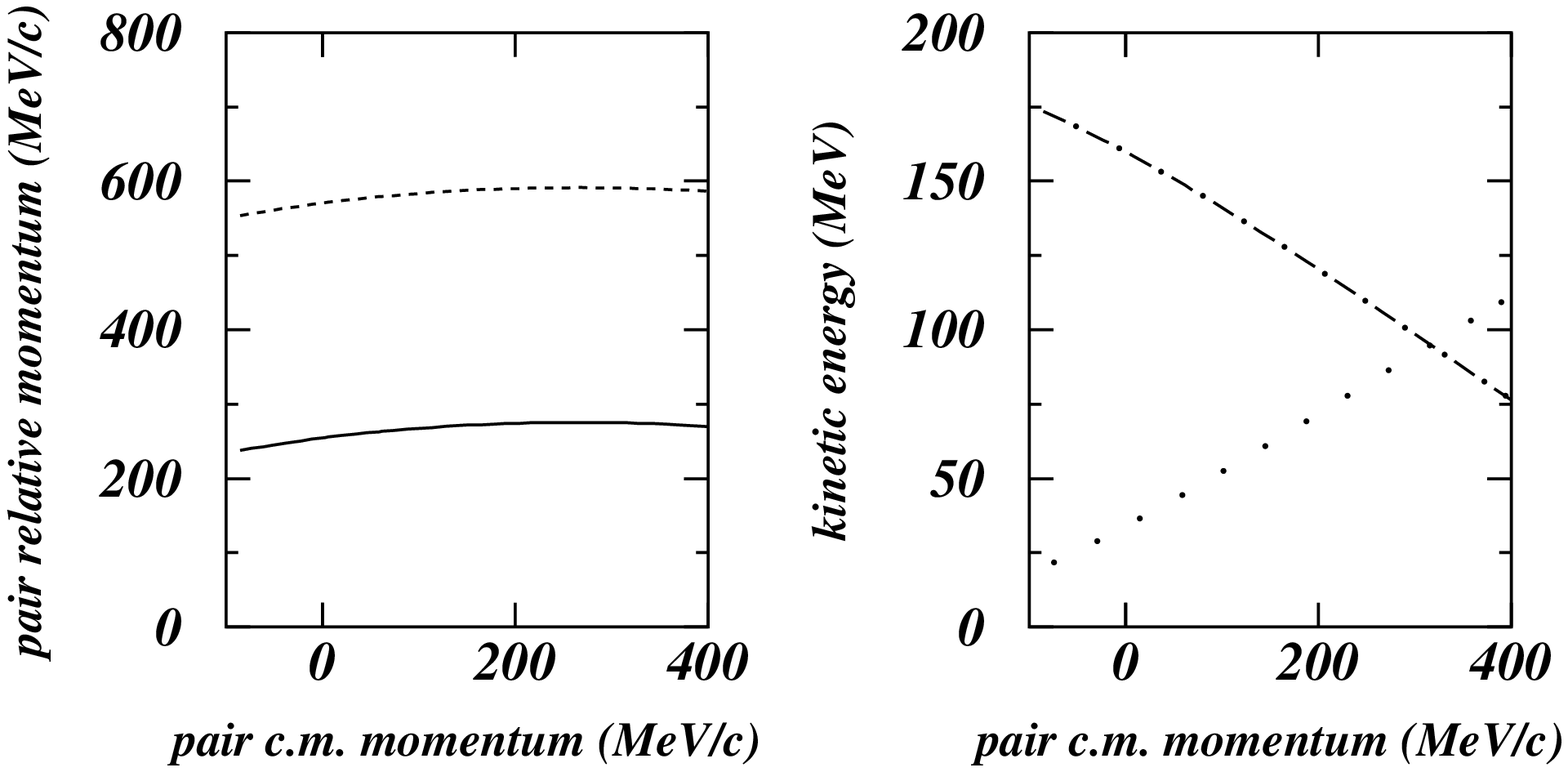}}
\end{center}
\caption{The kinetic energies (right panel) and pair relative momenta
$p_{rel} = \mid \frac {\vec{k_1} - \vec{k_2}} {2} \pm \frac {\vec{q}}
{2} \mid$ (left panel) as a function of the pair c.m. momentum for the
$^{16}$O$(e,e'pp)^{14}$C$(0^+,g.s.)$ reaction in super-parallel
kinematics at $<\omega>$=215~MeV and $<q>$=316~MeV. In the left panel,
the dashed (solid) line refers to the ``+'' (``-'') case of
Eq.~\ref{eq:relative}. In the right panel, the dot-dashed (dotted)
curve refers to the nucleon parallel (anti-parallel) to the momentum
transfer $\vec{q}$.}
\label{fig:kinemat}
\end{figure}

The first high-resolution $A(e,e'pp)$ data which could clearly separate
the individual states in the final nucleus became recently available
\cite{guenther2000,kahrau1999}.  The data were collected by the A1
collaboration with a unique three-spectrometer setup at the MAMI
facility in Mainz \cite{a1web}. An initial electron beam energy of
855~MeV and an $^{16}$O target was used.  The two ejected protons,
with momenta $\vec{k}_1$ and $\vec{k}_2$, were detected parallel and
anti-parallel to the momentum transfer, a peculiar situation which is
known as ``super-parallel kinematics''.  The energy and momentum
transfer was kept constant at a central value of $<\omega>$=215~MeV
and $<q>$=316~MeV.  Data were collected in a pair missing momentum $P
\equiv \mid \vec{k}_1 + \vec{k}_2 - \vec{q} \mid $ range of $-100 \leq
P \leq 400$~MeV/c.  In a naive spectator model the quantity $P$
corresponds with the center-of-mass (c.m) momentum of the diproton at
the instant of its interaction with the virtual photon.    The
quantity $p_{rel}$ is defined according to
\begin{equation}
p_{rel} = \left| \frac { \vec{k} _1 - \vec{k}_2 } {2} \pm \frac {\vec{q}} {2}
\right| \; .
\label{eq:relative}
\end{equation}
In an ideal world with vanishing final-state interactions, $p_{rel}$
would denote the relative momentum of the active proton pair before
its interaction with the photon, with the $-$ ($+$) sign in
Eq.~(\ref{eq:relative}) referring to the situation whereby the virtual
photon hits proton ``1'' (``2'').  Figure \ref{fig:kinemat} displays
the range in kinetic energies and pair relative momenta which is
covered in the $^{16}$O$(e,e'pp)^{14}$C experiment of
Refs.~\cite{guenther2000,kahrau1999} for which calculations will be
presented in this paper.  The curves refer to the kinematics
corresponding with the ground-state transition.  As we will restrict
ourselves to relatively low excitation energies in $^{14}$C, the
variation in kinetic energies and relative pair momenta is similar for
all transitions which will be considered here.  A strong variation in
the proton kinetic energies with the pair c.m. momentum is observed,
whereas the pair relative momentum is fairly constant.

\section{The model for electro-induced two-proton knockout}

The numerical $^{16}$O$(e,e'pp)$ calculations presented here, are
performed in a non-relativistic distorted-wave model outlined in
Refs.~\cite{jana568,jana624}. It is based on a partial-wave expansion
for the A-nucleon final state in terms of two-particle two-hole
($2p-2h$) eigenstates of the Hartree-Fock Hamiltonian.  A similar
approach has been adopted in Ref.~\cite{anguiano2003}. In this work,
the one-body Hartree-Fock Hamiltonian is derived self-consistently
starting from an effective nucleon-nucleon force of the Skyrme type.
The adopted partial-wave expansion for the final state reads
\begin{eqnarray}
\mid \Psi_f > & \equiv & \mid \Psi_{f}^{(A-2)}(E_x,J_R M_R);{\vec{k}}_1
m_{s_{1}} ; {\vec{k}}_2 m_{s_{2}} > \nonumber \\
& = & \sum_{l_{1}m_{l_{1}}j_{1}m_{1}}
      \sum_{l_{2}m_{l_{2}}j_{2}m_{2}}
(4 \pi)^2 i^{l_{1} + l_{2} }  
\frac {\pi} {2 m_p \sqrt{k_1 k_2}} 
\nonumber \\ 
& \times & e^{i(\delta_{l_{1}} +\sigma_ {l_{1}} 
+ \delta_{l_{2}}+\sigma_{l_{2}})}
\times Y_{l_{1} m_{l_{1}}}^{*}(\Omega_{1}) 
       Y_{l_{2} m_{l_{2}}}^{*}(\Omega_{2}) 
\nonumber \\ & \times & 
 <l_1 m_{l_{1}} \frac{1}{2}m_{s_{1}} \mid j_1 m_1> 
 <l_2 m_{l_{2}} \frac{1}{2}m_{s_{2}} \mid j_2 m_2> 
\nonumber \\
& \times  & 
\left| \Psi_{f}^{(A-2)}(E_x,J_R M_R) \right> \nonumber \\
& \times & 
\left| (p_1 (E_1 l_1 j_1 m_1) \; p_2 (E_2 l_2 j_2 m_2)) \right>
\; ,
\label{eq:wave}
\end{eqnarray}
where $m_p$ is the proton mass and the particle (or, continuum)
eigenstates $p_i$ of the mean-field Hamiltonian are characterized by
the quantum numbers $(E_i l_i j_i m_i)$. The energy $E_i$ of the
ejectile is determined by its momentum $\vec{k}_i$.  The central and
Coulomb phase shifts for the protons 1 and 2, are denoted by ($\delta
_{l_{1}}$, $\sigma _{l_{1}}$) and ($\delta _{l_{2}}$, $\sigma
_{l_{2}}$).  Further, $\left| \Psi_{f}^{(A-2)}(E_x,J_R M_R) \right>$
specifies the quantum numbers of the state in which the residual $A-2$
nucleus is left.  In our approach, the initial and final $A$-nucleon
states are orthogonal and anti-symmetrized.  The first property
implies that the overlap between the wave function for the
ground-state of the target nucleus and the wave functions of
Eq.~(\ref{eq:wave}) vanishes exactly. This is of particular importance
in view of the fact that in computing cross sections for triple
coincidence reactions characterized by small cross sections, great
care must be exercised to avoid all possible sources of spurious
contributions entering the overlap matrix elements.

As will become clear in the forthcoming discussions, an important
ingredient of the calculations are the two-hole overlap amplitudes
$X_{hh'}^{E_x}$ which determine the Two-nucleon Overlap Functions
(TOF) between the ground state of the target nucleus and each of the
probed states in the residual nucleus.  These coefficients play an
analogous role as the ``spectroscopic factors'' in $A(e,e'p)$
processes and are frequently referred to as $n \rightarrow n-2$ cfp
coefficients.  At this point, it is worth stressing that in the
analysis of single-proton knockout reactions of the $A(e,e'p)$ type,
the spectroscopic factors are usually treated as parameters which are
extraced by normalizing the computed to the measured differential
cross sections.

In a direct two-proton knockout process, solely the ``two-hole''
components in the final state will be excited.  Accordingly, that
piece of the final $A-2$ wave function which can be excited in a direct
two-proton knockout reaction can be written in terms of an expansion
of the form
\begin{eqnarray}
\left| \Psi_{f}^{(A-2)}(E_x,J_R M_R) \right> 
& = & \sum _{hh'} X _{hh'} ^{E_x} \left| \left( h^{-1}h'^{-1} \right) J_R \; M_R \right>
\nonumber \\   & = &
\sum _{hh'} X _{hh'} ^{E_x}
\sum_{m_hm_{h'}} \frac{1}{\sqrt{1+\delta_{hh'}}}
\nonumber \\   & \times &
\left< j_h m_h j_{h'} m_{h'} \mid J_R M_R \right> 
\nonumber \\
& \times & 
(-1)^{j_h+m_h+j_{h'}+m_{h'}}
\nonumber \\
& \times & 
c_{h-m_{h}}
c_{h'-m_{h'}} \left| \Psi_0 \right> \;,
\label{eq:exptwohole}
\end{eqnarray}
where $\left| \Psi_0 \right>$ is the ground-state of the target
nucleus which serves as the natural choice for the reference state.
In the $^{16}$O$(e,e'pp)$ reaction model calculations presented here,
the overlap amplitudes $X _{hh'} ^{E_x}$ are treated as input
parameters.  In determining their magnitudes we will be guided by
empirical information gathered in transfer reactions with hadronic
probes.  An alternative and ambitious approach, adopted e.g. by the
Pavia group, is to use the TOF (or, equivalently, the spectroscopic
information) predicted by advanced shell-model calculations
\cite{kadrev2003,giusti1998}.  Despite the enormous efforts directed
at calculating the TOF's, it turns out that the most sophisticated
two-hole spectral function calculations for $^{16}$O still miss some
key features of the low-lying states in $^{14}$C \cite{barbieri2002}.

Qualitatively, the missing energy spectra extracted from the analysis
of the Mainz high-resolution $^{16}$O$(e,e'pp)$ experiment bears a
strong resemblance with those obtained in a $^{15}$N$(d,^3He)^{14}$C
\cite{kaschl71} and $^{16}$O$(^{6}$Li$,^8$B$)^{14}$C
\cite{weisenmiller} measurement.  In essence, the ground-state
($0^+$), a doublet of $2^+$ states (respectively, at $E_x$=7.01 and
8.32~MeV), a $1^+$ state at $E_x$=11.31~MeV and a $0^+$ state at
$E_x$=9.75~MeV are populated.  As a matter of fact, we will exploit
this similarity to guide our choices with respect to the spectroscopic
information for the two-nucleon overlap functions entering the
numerical calculations.

In essence, the eight-fold $(e,e'pp)$ differential cross sections are
computed starting from the transition amplitude
\begin{equation}
J^{\mu}(\vec{q}) = \int d \vec{r} \left< \Psi_{f} \left| e ^ {i
  \vec{q} \cdot \vec{r} } J ^{\mu} \left( \vec{r} \right) \right|
  \Psi_{0} \right> \; , 
\end{equation}
where $\vec{q}$ is the momentum-transfer induced by the virtual
photon, and $J^{\mu}$ the nuclear current operator.  The operator
$J^{\mu}$ considered here is the sum of the $\Delta$-isobar current
operator, and an operator which is the product of the one-body charge-
and current-density $J^{\mu}_{[1]}(\vec{r})$ and a central correlation
function $g(r_{12} \equiv \mid \vec{r}_1 - \vec{r}_2 \mid)$
\begin{eqnarray}
& & \biggl( J^{\mu} _{[1]} (\vec{r}_1) + J^{\mu}_{[1]} (\vec{r}_2) \biggr)
g \left( r_{12} \right) \; + \nonumber \\ 
& & g ^{\dagger} \left( r_{12} \right) 
\biggl( J^{\mu} _{[1]} (\vec{r}_1) + J^{\mu}_{[1]} (\vec{r}_2) \biggr)
\; .
\label{eq:srccurrent}
\end{eqnarray}
In this current operator, the $g(r_{12})$ term accounts for the effect
of SRC in the pair wave function at the time that it was hit by the
virtual photon and belongs to the class of initial-state correlations.
The $g ^{\dagger} (r_{12})$ term, on the other hand, is part of the
class of final-state correlations and implements the effect of SRC on
the wave function for the two ejected protons. In order to guarantee
the hermiticity of the above current operator (\ref{eq:srccurrent}),
the inital- and final-state short-range correlations are modeled with
the aid of the same correlation function.  The central correlation
function $g(r_{12})$ expresses how strongly two nucleons, which are
$r_{12}$ apart, are correlated. In the absence of nucleon-nucleon
correlations beyond those already implemented in the
independent-nucleon picture, this contribution would simply vanish. A
plethora of parameterizations for $g(r)$ can be found in the
literature. They range from those that have a hard core extending over
more than 1 fm to those with a relatively soft core and sizable
probability to find two protons at the same position. Our numerical
results are obtained with the correlation function of Ref.~\cite{GH}
which categorizes somewhere in between these two extreme classes.  The
correlation function of Ref.~\cite{GH} was earlier found to provide a
favorable agreement with the $^{12}$C$(e,e'pp)$ measurements of
Ref.~\cite{raoul} and the $^{16}$O$(e,e'pp)$ data reported in
Ref.~\cite{starink}.  The $\Delta$-current operator used in this work
has an energy- and medium-dependent $\Delta$ width.  Its detailed form
can be found in Ref. \cite{janeep}.

\begin{table}
\begin{tabular} {|l||c|c|c|c|} \hline
&  Ref.~\cite{geurts96} & Ref.~\cite{cohen70} & Ref.~\cite{ensslin74}
& Ref.~\cite{huffman} \\ \hline \hline
m & 0.97 & 0.91 & 0.76 & 0.576 \\
n & 0.24 & 0.41 &-0.65 & 0.818 \\ \hline
\end{tabular}
\caption{Two-hole overlap amplitudes for the $^{16}$O$\rightarrow
  ^{14}$C ground-state transition.}
\label{tab:tab1}
\end{table}

\section{Results and Discussion}

\begin{figure}
\begin{center}
\resizebox{0.48\textwidth}{!}{\includegraphics{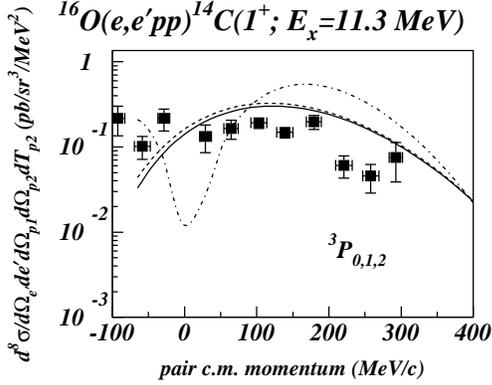}}
\end{center}
\caption{The eightfold differential cross section for the
$^{16}$O$(e,e'pp)^{14}$C$(1^+,E_x=11.31~MeV)$ reaction as a function
of the pair c.m.  momentum. The dashed curve shows the results of
the distorted-wave calculations that include only intermediate
$\Delta$ excitation.  The solid (dot-dashed) curve is the result of a
distorted-wave (plane-wave) calculation that accounts for both
intermediate $\Delta$ and central short-range correlations. The data are from
Refs.\cite{guenther2000} and \cite{kahrau1999}.}
\label{fig:fin1p}
\end{figure}

We start our discussion of the $^{16}$O$(e,e'pp)$ differential cross
sections with the $1^+$ state at $E_x=$11.31~MeV.  In most
nuclear-structure calculations, the two-proton overlap amplitudes for
this particular state are dominated by the $\left| \left( 1p_{3/2}
\right) ^{-1} \left( 1p_{1/2} \right) ^{-1} ; 1^+ \right>$ two-hole
configuration.  As we use realistic single-particle wave functions
obtained through solving the Hartree-Fock Hamiltonian, an exact
separation into the relative $\vec{r}_{12} = \vec{r} _1 - \vec{r}_2 $
and c.m. $ \vec{R} = \frac {\vec{r}_1 + \vec{r} _2} {2} $ coordinate
is not possible in our approach. The well-known Moshinsky
transformation for a harmonic-oscillator basis can however serve as a
guide to identify the dominant relative and c.m. quantum numbers of
the diprotons for a specific transition.  Indeed, the quantum numbers
of the final state impose strong restrictions on the possible
combinations for the relative and c.m. angular momentum of the active
diproton. For the $\left| (1p)^{-2} ; 1^+ \right>$ configuration only
the combination of $L=1$ c.m. and $P$-wave relative wave functions is
allowed.  In what follows, we will denote the c.m. angular momentum
with $L$.

The plane-wave $^{16}$O$(e,e'pp)^{14}$C($E_x=$11.31~MeV) predictions shown
in Fig.~\ref{fig:fin1p} clearly exhibit this $L=1$ behavior.  The
distortions which the struck protons undergo through the presence of
the other target nucleons, fill in the dip in the $P$-wave pair
c.m. momentum distribution about $P \approx 0$.  This peculiar feature
is also observed in the data.  A striking feature of the calculations
displayed in Fig.~\ref{fig:fin1p} is that the $\Delta$ contribution is
by far the dominant one, while central short-range correlations are
only marginally contributing.  The distorted-wave model provides a
reasonable description of the data.  The curves displayed in
Fig.~\ref{fig:fin1p} use an overlap amplitude
$X_{hh'}^{E_x=11.31~MeV}$ of 1 for the sole component $\left| \left(
1p_{3/2} \right) ^{-1} \left( 1p_{1/2} \right) ^{-1} ; 1^+ \right>$.
This number is about 30\% larger than the value of 0.76 quoted in
Ref.\cite{geurts96}.

\begin{figure}
\begin{center}
\resizebox{0.48\textwidth}{!}{\includegraphics{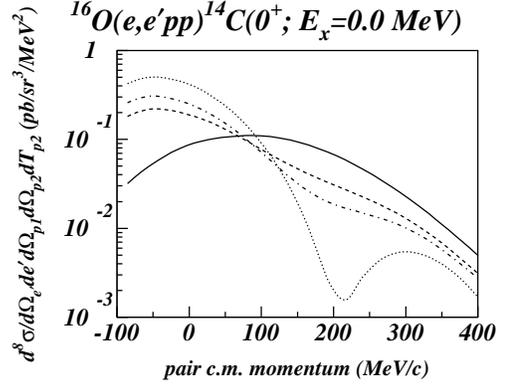}}
\end{center}
\caption{Calculated differential cross sections for the
$^{16}$O$(e,e'pp)^{14}$C$(0^+,E_x=0~MeV)$ reaction using various sets
of two-proton overlap amplitudes.  All curves are obtained in a
distorted wave approximation and account for central short-range
correlations and intermediate $\Delta$ excitation.  The calculations
use the two-proton overlap amplitudes $(m,n)$ from
Ref.~\protect \cite{ensslin74} (solid curve), Ref.~\protect \cite{cohen70} (dot-dashed
curve), Ref.~\protect \cite{geurts96} (dashed curve) and Ref.~\protect
\cite{huffman}
(dotted curve). The corresponding values for $(m,n)$ are listed in
Table~\ref{tab:tab1}. }
\label{fig:gsstruc}
\end{figure}

\begin{figure}
\resizebox{0.48\textwidth}{!}{\includegraphics{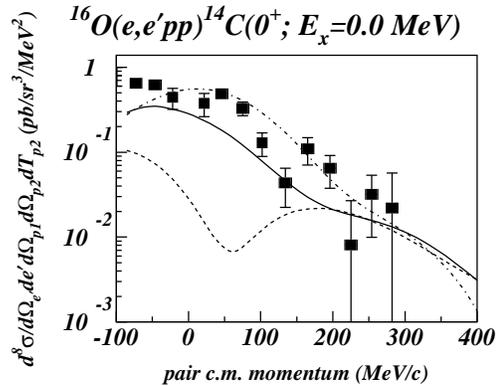}}
\caption{The eightfold differential cross section for the
$^{16}$O$(e,e'pp)^{14}$C$(0^+,E_x=0~MeV)$ reaction as a function of
the pair c.m. momentum. The dashed curve shows the results of the
distorted-wave calculations that include only intermediate $\Delta$
excitation.  The solid (dot-dashed) curve is the result of a
distorted-wave (plane-wave) calculation that accounts for both
intermediate $\Delta$ and central short-range correlations. The data
are from Refs.\cite{guenther2000} and \cite{kahrau1999}.}
\label{fig:fin0p}
\end{figure}

We proceed with discussing the results for the ground-state
transition.  In contrast to the situation for the $1^+$ state, at
least two two-hole configurations have been frequently quoted as major
contributors to a two-proton transfer process.  As a matter of fact,
in the spirit of Eq.~\ref{eq:exptwohole} one can write that the
relevant components in the ground state of $^{14}$C are
\begin{eqnarray}
\left| 0^+ ; E_x = 0~MeV \right > & = &  m \left| \left( 1p_{1/2} \right)
^{-2}  ; 0^+ \right> \nonumber \\ 
& & + n \left| \left( 1p_{3/2} \right)
^{-2} ; 0^+ \right> \; .
\end{eqnarray}
The precise values of $m$ and $n$ are subject to discussion, though.
Table~\ref{tab:tab1} summarizes some of the combinations of values for
the two-hole overlap
amplitudes $m$ and $n$ that can be found in literature.  Additional
sets can be found in Tables II and III of Ref.~\cite{huffman}.  For
the sake of clarity the overlap amplitudes contained in
Table~\ref{tab:tab1} were normalized by putting $m^2+n^2=1$.

The sensitivity of the computed
$^{16}$O$(e,e'pp)^{14}$C$(J^{\pi}=0^+,E_x=0~MeV)$ differential cross
sections to the choices made with respect to the magnitude of the
overlap amplitudes is displayed in Fig.~\ref{fig:gsstruc} containing
predictions for each set contained in Table~\ref{tab:tab1}.  The solid
line, which corresponds with overlap amplitudes that are obtained
from fitting inelastic $M1$ form factors for the
$^{14}$N$(e,e')(J^{\pi}=0^+, T=1, E_x=2.313~MeV)$ transition
\cite{ensslin74}, exhibits a c.m. momentum dependence which is completely out
of line from the other three curves. The $\left| \left( 1p \right)
^{-2} ; 0^+ \right>$ configuration gives rise to relative and c.m. wave
functions corresponding with relative $^1S_0$ in combination with
$L=0$ and relative $^3P_1$ combined with $L=1$.

The results of the NIKHEF $^{16}$O$(e,e'pp)^{14}$C($0^+$)
experiments \cite{gerco1998,starink} provide strong evidence for the
presence of a strong $L=0$ component, thereby excluding the solid curve
in Fig.~\ref{fig:gsstruc}.  It is worth remarking that the overlap
amplitudes corresponding with the solid curve stem from a fit to the
$M1$ inelastic $^{14}$N$(e,e')$ form factor that is not particularly
sensitive to the $L=0$ component.  The other three curves in
Figure~\ref{fig:gsstruc} correspond with a relative contribution of
the $(1p_{1/2})^{-2}$ and $(1p_{3/2})^{-2}$ components that is
gradually changing.  The larger the contribution from
$(1p_{3/2})^{-2}$, the larger the $L=0$ component and the smaller the
$L=1$ component.

Figure \ref{fig:fin0p} displays a comparison of the recently obtained
$^{16}$O$(e,e'pp)^{14}$C$(0^+,E_x=0~MeV)$ data and our reaction model
calculations using the two-hole overlap amplitudes of
Ref.~\cite{cohen70}.  The distorted-wave calculations including
short-range correlations reproduce the missing-momentum dependence
well, while underestimating the data by roughly a factor of two over
the whole momentum range.  We wish to stress that with the two-hole
overlap amplitudes $(m,n)$ from Ref.~\cite{cohen70}, the presented
model produced also a reasonable agreement with the
$^{16}$O$(e,e'pp)^{14}$C$(0^+,E_x=0~MeV)$ cross sections measured at
the AMPS facility \cite{starink}. These overlap amplitudes compared
also favorably with the $^{15}$N$(d,^3He)^{14}$C$(0^+,g.s.)$
measurements reported in Ref.~ \cite{kaschl71}.  An interesting
observation from Fig.~\ref{fig:fin0p} is that the distorted-wave
calculation ignoring central short-range correlations, underestimates
the data at low pair missing momenta by several factors.  At high pair
missing momenta, where the $L=1$-wave can be expected to dominate, the
calculations neglecting the central short-range correlations move
closer to the data.  In any case, without inclusion of central
short-range correlations, neither the shape nor the magnitude of the
data for the ground-state transition can be reproduced. We interpret
this as strong evidence for short-range correlations for proton pairs
residing in relative $^1S_0$ states.  At the same time, and equally
important, central short-range correlations appear to affect
exclusively proton pairs in relative $S$ and c.m. $L=0$ states.

\begin{figure}
\begin{center}
\resizebox{0.48\textwidth}{!}{\includegraphics{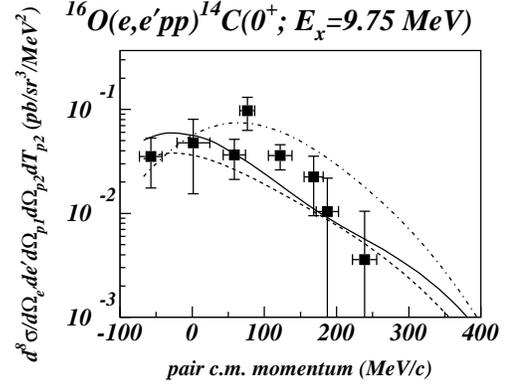}}
\end{center}
\caption{As in Figure \protect \ref{fig:fin0p} but now for the
excitation of the $0^+$ state at an excitation energy of 9.75~MeV. The
data are from Refs.\cite{guenther2000} and \cite{kahrau1999}.}
\label{fig:fins0p}
\end{figure}

The superior resolution at the unique MAMI three-spectrometer setup
made determining the differential cross sections for a weakly excited
$0^+$ state at $E_x=$9.75~MeV feasible.  It is tempting to interpret
this state as the ``orthogonal'' partner of the ground-state.
Shell-model calculations, however, predict that the ``orthogonal
partner'' of the ground state is located at a substantially larger
excitation energy (respectively, $E_x$=16.32~MeV in the calculations
of Ref.~\cite{cohen70} and 12.00~MeV in the calculations of
Ref.~\cite{geurts96}).  Using the overlap amplitudes quoted in
Ref.~\cite{cohen70} we obtain a differential cross section which
overshoots the data by several factors over the $P$ range.  This
suggests that the 9.75~MeV state is not the ``orthogonal'' partner of
the ground-state and can at most carry a fraction of the corresponding
two-hole strength.  Indeed, assuming that about 17\% of the two-hole
strength contained in the wave function of Ref.~\cite{cohen70}
\begin{eqnarray}
\left| 0^+ _2 \right > & = & + 0.41 \left| \left( 1p_{1/2} \right)
^{-2}  ; 0^+ \right> 
\nonumber \\
& & - 0.91 \left| \left( 1p_{3/2} \right)
^{-2} ; 0^+ \right> \; ,
\end{eqnarray}
is carried by the 9.75~MeV state, we obtain the results displayed in
Figure~\ref{fig:fins0p}.  It is worth stressing that also the
$^{15}$N$(d,^3He)^{14}$C measurements reported in Ref.~\cite{kaschl71}
do not provide evidence for a sizable population  of a $0^+$ state
for $E_x \approx 10$~MeV.

\begin{figure}
\begin{center}
\resizebox{0.48\textwidth}{!}{\includegraphics{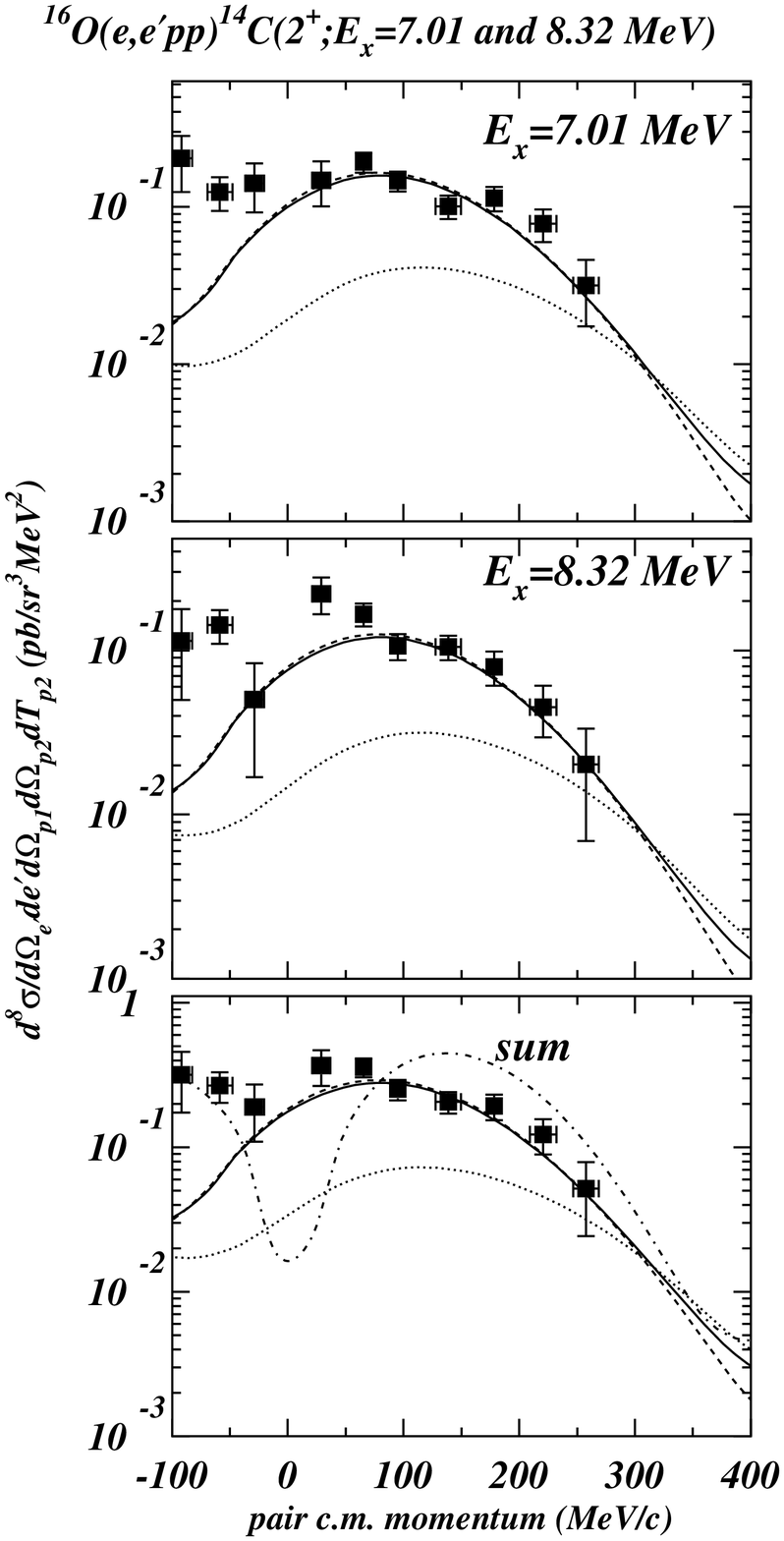}}
\end{center}
\caption{The eightfold differential cross section for the
$^{16}$O$(e,e'pp)^{14}$C$(2^+,E_x$=7.01 and 8.32~MeV) reaction as a
function of the pair c.m. momentum. The dashed curve shows the
results of the distorted-wave calculations that include only
intermediate $\Delta$ excitation.  The solid (dot-dashed) curve is the
result of a distorted-wave (plane-wave) calculation that accounts for
both intermediate $\Delta$ and central short-range correlations. The
dotted line omits the core excitations of the $(1d)^2$ type from the
full calculations.The data are from Refs.\cite{guenther2000} and
\cite{kahrau1999}.}
\label{fig:fin2p}
\end{figure}

The nuclear-structure calculations of Refs. \cite{cohen70} and
\cite{geurts96} both assign the following structure
\begin{eqnarray}
\left| 2^+ \right> & = & + 0.976 \left| \left( 1p_{3/2} \right)
^{-1}  \left( 1p_{1/2} \right) ^{-1}  ; 2^+ \right> \nonumber \\ 
& & +
0.212 \left| \left( 1p_{3/2} \right)
^{-2}  ; 2^+ \right> \; , 
\label{eq:wav2p}
\end{eqnarray}
to the lowest 2$^+$ state in $^{14}$C.  In the analysis of the
transfer reaction $^{15}$N$(d,^3He)^{14}$C measurement reported in Ref.~\cite{kaschl71}, it was found
that the low-lying $2^+$ strength is fragmented over at least three
states.  In the same paper, a careful analysis of the data led to the
conclusion that {\em ``whereas the $J^{\pi}$=0$^+$ and 1$^+$,T=1 states
of mass 14 are rather pure $(1p)^{-2}$ states, the 2$^+$,T=1 states
are STRONGLY mixed with core excitations of the $(2s_{1/2},1d)^2$
type''}.

The dotted curves in Figure \ref{fig:fin2p} are obtained with the
overlap amplitudes contained in Eq.~(\ref{eq:wav2p}).
For the fragmentation of the strength over the different physical
$2^+$ states we adopt the values as they were obtained in the
aforementioned analysis of $^{15}$N$(d,^3He)^{14}$C measurements
(i.e., 47\% and 36\% for the 7.01 and 8.32~MeV state in the doublet).
It is clear that when including solely $(1p)^{-2}$ two-hole overlap
amplitudes our distorted-wave calculations badly fail in predicting
the shape and magnitude of $^{16}$O$(e,e'pp)^{14}$C differential cross
sections.  A similar result has been obtained with the Pavia
$A(e,e'pp)$ model \cite{guenther2000}. These failures are probably not
so surprising in the light of the findings for the $2^+$ transitions
in $^{15}$N$(d,^3He)^{14}$C measurements \cite{kaschl71}.

We have taken up the aforementioned suggestion based on an analysis of
$^{15}$N$(d,^3He)^{14}$C hadron reactions, that important core
components may contribute for the $2^+$ excitation. After including
core polarizations of the $(1s)^2(1p)^4(1d)^2$ type in the target
ground-state wave function, which amounts to modifying the $\left| \Psi_0 \right>$ wave
functions of Eq.~(\ref{eq:exptwohole}) into
\begin{eqnarray}
\left| \Psi_0 \right> & = & \sqrt{1 - \beta ^2} \left| (1s)^2 \; (1p)^6
\right> \nonumber \\
& + & \beta  \left| (1s)^2 \; (1p)^4 (1d_{5/2})^2 \right>
\end{eqnarray}
we find the solid curves in Figure~\ref{fig:fin2p}.  The least one can
say is that core polarizations have a large impact on the calculated
cross sections for electro-induced two-proton knockout to the $2^+$
state. After including the core polarization effects, a reasonable
description of the data is reached.  For the curves of
Figure~\ref{fig:fin2p}, the $(1s)^2(1p)^4(1d)^2$ core polarization was
implemented with an amplitude of $\beta = 0.4$ which is a value
suggested by the shell-model calculations of Ref.~\cite{true}.

Our predictions for the cross sections tend to systematically
underestimate the data at negative values of the c.m. momentum $P$.
As can be appreciated from inspecting the right panel of
Fig.~\ref{fig:kinemat}, negative values of $P$ correspond with an
asymmetric situation with an extremely slow foreword going proton.
For these slow moving protons, it cannot be excluded that
multiple-scattering contributions, not included in the presented
calculation, gain in relative importance.

\section{Summary}
We have computed the eightfold $^{16}$O$(e,e'pp)$ differential cross
sections in a distorted-wave model adopting a direct reaction
process. Two-nucleon photoabsorption mechanisms involving intermediate
$\Delta$ creation and central short-range correlations have
been implemented.  In non-relativistic approaches, as the one adopted
here, meson-exchange currents do not contribute to two-proton knockout
reactions. In general, the $\Delta$-isobar current is at the origin of
the major fraction of the electroinduced two-proton knockout strength,
thereby confirming the conclusions drawn in earlier investigations
\cite{raoul}.  The effect of central short-range correlations becomes
clearly visible for the peculiar case that a diproton remaining in a
relative $S$ state can additionally be guaranteed to have a small
c.m. momentum.  Such conditions can be studied in the $^{16}$O$(g.s.)
(e,e'pp) ^{14}$C$(g.s.)$ reaction.  At high c.m. momenta, the
differential cross section for this transition is indeed dominated by
the $\Delta$-isobar current, the short-range correlations providing
marginal amounts of strength.  At low c.m. momenta, the opposite is
true. In line with the observations made in hadronic transfer
reactions, a proper description of the excitation of the $2^+$ states
in $^{14}$C requires a strong mixing of the $^{16}$O ground-state with
long-range core polarization components.  The presented investigations
illustrate that besides questions related to the description of
final-state interaction effects and the implementation of two-body
currents, an analysis of $A(e,e'pp)$ reactions is highly sensitive to
the spectroscopic information.  Indeed, unlike in the exclusive
$A(e,e'p)$ case, where one can usually identify one dominant
single-hole component for a particular transition, the two-nucleon
knockout process to individual states is often the result of several
strongly interfering two-hole overlap amplitudes.  For the latter
quantities, which refer to the long-range dynamics of nuclei, widely
varying predictions can be found in literature. With respect to the
magnitudes for the two-hole overlap amplitudes, we adopt a heuristic
view and found that the values which did fairly well in explaining
hadron transfer reactions, also provide a reasonable description of
the $^{16}$O$(e,e'pp)$ angular cross sections to individual states.
Often, a consistent description of the short-range (central
correlation function) and the long-range (two-hole overlap amplitudes)
dynamics of nuclei is considered essential to arrive at a rigorous,
coherent theoretical picture of the $A(e,e'pp)$ reaction.  The effect
of SRC being confined to proton pairs in a relative $S$ state, the
Pauli principle decouples the effect of short-range and long-range
dynamics to a large degree.  Indeed, for the results presented here
the expected synergy between the short- and long-range dynamics is
solely applicable to the $^{16}$O$(g.s.)(e,e'pp) ^{14}$C$(g.s.)$
reaction at low pair c.m. momenta.  The $^{16}$O$(g.s.)(e,e'pp)
^{14}$C$(g.s.)$ reaction at high pair c.m. momentum and the
transitions to the other low-lying states in $^{14}$C are rather
dominated by the synergy of $\Delta$ degrees-of-freedom and the
long-range dynamics of the nuclei involved.


\end{document}